\documentclass[a4paper,12pt,twoside]{cpc-hepnp}
\usepackage{multicol}
\usepackage{graphicx}
\usepackage{booktabs}
\usepackage{lscape}
\usepackage{rotate}
\usepackage{amssymb,bm,mathrsfs,bbm,amscd}
\usepackage[tbtags]{amsmath}
\usepackage{lastpage,makecell}

\begin{document}

%\begin{CJK*}{GB}{gbsn}
%\begin{CJK*}{GBK}{song}

%\fancyhead[c]{\small Chinese Physics C~~~Vol. xx, No. x (201x) xxxxxx}
%\fancyfoot[C]{\small 010201-\thepage}

%\footnotetext[0]{Received 31 June 2015}

\title{The Ground State Aspects and the Impact of Shell Structures on the Stability of Es-Isotopes }

\author{%
\quad C. Dash$^{1*}$,\email{anuchinu20@gmail.com}
\quad A. Anupam$^{2}$,
\quad I. Naik$^{1}$,
\quad B. K. Sharma$^{3}$, and
\quad B. B. Sahu$^{2*}$\email{bbsahufpy@kiit.ac.in}
}
\maketitle

\address{%
$^1$ Department of physics, MSCBD University, Baripada 757003, India\\
$^1$ Department of physics, Anchalika Mahavidyalaya Gadia, Mayurbhanj-757023, Odisha, India\\
$^2$ School of Applied Sciences, KIIT Deemed to be University, 
Bhubaneswar-751024, Odisha, India\\
$^3$ Department of Physics, Amrita School of Physical Sciences, Amrita Vishwa Vidyapeetham, Coimbatore-641112, India\\
}

\begin{abstract}

 In this work, we have analyzed the nuclear structure and several prospective decay characteristics of the $^{240-259}$Es$_{99}$ isotopes.
For this we use Relativistic Mean Field model (RMF) with NL-SH and NL3* force parameter in an axially deformed oscillator basis.
  In structural properties, we have analyzed binding energy (B.E.), skin thickness ($r_{np}$) , charge radius ($r_c$), one neutron separation energy ($S_{1n}$), 
two neutron separation energy ($S_{2n}$), differential variation of two neutron separation energy ($dS_{2n}$), the single particle energy and its variation with quadrupole deformation parameter of Es isotopes.
 We have also estimated the $\alpha$-decay, $\beta$-decay and cluster decay half lives of Es isotopes to analyze the shell structure and also to predict the suitable decay mode among them. 
 The $\alpha$-decay half-life periods are calculated using the MUDL and AKRE formulae using both our calculated Q-values and empirically assessable Q-values.
In a similar manner, we have computed the half-lives of cluster decay using Universal Decay Law and HOROI formula. A longer decay half-life indicates a shell stabilized parent nucleus, while a small parent half-life suggests the shell stability of the daughter. This study provides us the insights regarding the structural changes with the change in neutron number enabling us to predict shell closures and nuclear stability. We found a shell/sub-shell closure at N = 154 for the NL-SH parameter set. 
This research aids in our comprehension of Es isotopes' shell structure and decay mechanism.

\end{abstract}


\begin{pacs}
{25.60.Bx; 25.60.Pj; 25.70.Ef}
\end{pacs}

%\footnotetext[0]{\hspace*{-3mm}\raisebox{0.3ex}{$\scriptstyle\copyright$}2013
%Chinese Physical Society and the Institute of High Energy Physics
%of the Chinese Academy of Sciences and the Institute
%of Modern Physics of the Chinese Academy of Sciences and IOP Publishing Ltd}%

\begin{multicols}{2}

\section{Introduction}
The stability of a nucleus lying especially in the actinide region of the periodic table depends on many factors such as a greater binding energy per nucleon (B.E./A), a proper N/P ratio, the closed shell structure etc. But the overall binding energy per nucleon of actinide nuclei is lower than that of medium-mass nuclei which in turn favors radioactivity in this region. Hence the role of closed shell effects associated with protons or neutrons becomes more important in providing stabilization to these nuclei causing an increase in nuclear lifetimes. The shell effect delays the decay process. As an example, a delay of about 15 orders of magnitude is found in the isotope of Fm with the deformed closed shell at N = 152 \cite{pat89}. 

 The height of the fission barrier determines the chances of spontaneous fission. Being very common and prominent in actinide region, nuclear deformation modifies the structure of single-particle energy levels influencing the fission barrier. 
There are many theoretical and experimental investigations carried out to find out the factors which helps in identifying the shell/sub-shell closures in the heavy and super heavy region. Among them alpha decay and cluster decay play a very important role in identifying the shell closure. A smaller decay life indicates the presence of spherical or deformed shell closure in the daughter nucleus whereas high value of decay life indicates a shell closure of parent nucleus \cite{kum09}. 

 Investigation of skin thickness \cite{vir19} and rms charge radius are also important towards the structural study of a nucleus as they are sensitive towards any change in size and shape of the nucleus. Some times we observe a prominent kink in $r_c$ across spherical
shell closures. \cite{war24}. 
 The separation energies also provide us informations regarding the structure of a nucleus and nature of shell/sub-shell closure.
Single particle energies and its variation with the quadrupole deformation along with these bulk properties give us the insights regarding structural evolution of an isotopic series.

In recent years, the shell gap at N = 152 beyond $^{208}Pb$ was predicted by various studies \cite{ram12, the15}. The gap at N = 152 was first experimentally found in $^{250}Cf$ isotope in 1954 by Ghiorso et al \cite{ghi54}. 
 During the structural investigation of Thorium isotopes based on nuclear
density functional theory, Ummukulsu \textit{et al} \cite{umm23} 
predicted a sub-shell closure at N = 138. 
Comparing the results of the ratio of quadrupole deformation to the pairing gap parameter with the simple factor (P), Brenner et al \cite{bre94} predicted a persistent spherical sub-shell gap at N=164. With a proper investigation of Nilsson diagram Gustafson et al \cite{gus67} also, suggested a spherical shell gap at N = 164 \cite{bre94}. 
From the study of single particle energy level for $^{256}U_{92}$ using KY potential Ishii et al \cite{isi07} found a clear shell gap at N = 164 for zero deformation.  
From $\alpha$-decay investigations Ismail et al \cite{ism10}
predicted enhanced stabilities at N = 152 and 162 predicting shell closures at these neutron numbers. We have also reproduced the shell closures at N = 152 and 162 for Fm \cite{das25}. 
Some theoretical models have also predicted that $^{270}Hs$ (N = 162) is the deformed doubly magic nucleus \cite{laz94, hof02}
 and it is the first experimentally observed even-even nucleus on the predicted realm of neutron shell closure of N = 162 \cite{dvo06}.

	 All the above mentioned studies encouraged us to probe the spherical or deformed shell/sub-shell closures above $^{208}Pb$. Here we choose Einsteinium and carried out the investigations regarding the structural properties, $\alpha$ decay, $\beta$-decay and cluster decay half lives of $^{240-259}$Es$_{99}$ isotopes.
	 
	 At first Einsteinium was discovered accidentally along with Fermium from the debris of the first thermonuclear weapon test at Eniwetok Atoll in 1952 \cite{yui89}. 
That is $^{253}$Es$_{99}$ isotope. 
It had been thought to be produced by neutron capture (15 neutrons) by $^{238}$U$_{92}$ nuclei followed by 7$\beta$ decays. 
Due to high neutron flux density during detonation such multiple neutron capture was possible. 
$^{253}$Es$_{99}$ was first synthesized by Thompson et al \cite{yui89} by multi neutron capture on $^{239}Pu$ in 1954. 
Now 19 isotopes of Es such as $^{240-257}$Es$_{99}$ and three nuclear isomers of Es are known to us. 
The element with the greatest atomic number that is found in macroscopic quantities in its pure form among synthetic elements is Es ($^{253}$Es$_{99}$) \cite{hai79}. 
 
Because of their comparatively greater cross-sections \cite{cho15}, the nuclei around Z = 100 act as the heaviest long lived radioactive targets for further synthesis of neutron rich nuclei in the super heavy element (SHE) region. Here we can see the applications of Es isotopes. 
In 1955 $^{256}$Md$_{101}$ was synthesized by using $^{253}$Es$_{99}$ as a target \cite{ghi16}. 
In 1985 $^{254}$Es$_{99}$ was used as a target for the synthesis of element Z = 119 by bombarding it with $^{48}$Ca$_{20}$ ions at super HILAC linear particle accelerator at Berkeley California but the attempt remained unsuccessful. 
No atoms were identified \cite{lou85}. 
Hence it is very important to have a detailed knowledge about the structural properties and decay modes of Es isotopes.

		  To calculate the structural properties we have employed the relativistic mean field model (RMF). The RMF model is one of the successful theoretical models in describing the ground state properties of nuclei over the periodic table \cite{gam90, lal97}. In earlier works it is successfully explained for different elements by our collaborators \cite{sah18, swa16, swa17, swa18, swai18, swain18, das19, swa19, das21}. The key features of RMF model behind its wide use are
	  
	  (i) The model consider a nucleus as an aggregation of Dirac nucleons and they interact with the exchange of mesons in a relativistically covariant manner.
	  
	  (ii) Instead of forces, the fields which are characterized by their angular momentum, parity and isospin mediate the interaction.
	  
	  (iii) The model naturally includes the spin orbit interaction
	  
As RMF model is a self-consistent parameter dependent model, we choose two non-linear force parameters that are NL3* force parameter \cite{lol09} and NL-SH force parameter \cite{sha93} during our calculations.

	In structural studies, we have computed the bulk properties such as binding energy (B.E.), binding energy per nucleon (B.E./A),quadrupole deformation parameter ($\beta_{2}$), separation energies, differential variation of two neutron separation energy (d$S_{2n}$) of  $^{240-259}$Es$_{99}$ isotopes. 
	We have studied the single particle energies for $^{247}Es$ and $^{257}Es$ isotopes and also studied the variation of single particle energies with deformation parameter by drawing the Nilson plot for $^{257}Es$ isotope. In addition, we have computed the $\alpha$-decay, $\beta$-decay and cluster decay half-lives. We have used two semi-empirical formulas, MUDL \cite{akra19} and AKRE \cite{akr19} for $\alpha$-decay calculations. Here we have calculated the $Q_{\beta}$-values following the article \cite{mol19}.
 We  Use both our calculated and experimentally available $Q_{\beta}$ -values a newly developed semi empirical formula from the article \cite{sob23} 
 to calculate $\beta$-decay half lives.  
 The cluster decay half-lives 
have also been computed by using the universal decay rule \cite{ism17}
and the cluster decay scaling law \cite{ade16}. The binding energy values obtained from the NL3* force parameter \cite{lol09} and the NL-SH force parameter \cite{sha93} are utilized to determine the Q-values for above mentioned decay processes. 
These studies helped us in deep examination of the structure of Es isotopes.

	Theoretical formulation is given in Section. 2. Sec. 3 contains the analysis of findings and Sec. 4 provides a summary of the results.

\label{intro}

\section{Theoretical framework}

In 1951, Schiff \cite{sch51} proposed the relativistic notion of a nuclear system, speculating that nuclear saturation may result from the strong nonlinear self interactions among scalar fields. Subsequently, the idea undergone periodic modifications by several theoretical nuclear physicists, including Teller and Durr in 1956 \cite{john55, dur56} and Green and Miller in 1972 \cite{gren72}. In 1974 Walecka and his collaborators developed the simplest version of relativistic quantum field model to deal with nuclear many body problem \cite{wal74, wal92} taking Lagrangian as their starting point. In this model the nucleons and mesons are considered as the degrees of freedom. They also assumed that the mesons do not interact with each other. This model is also known as $\sigma -\omega$ model. The model was again modified by Boguta and Bodmer \cite{bog77} with an addition of non linear self-coupling terms of $\sigma$-field to the above Lagrangian. By including nonlinear self coupling components of the $\omega$ meson in the Lagrangian, the RMF model is further altered. The source code used here is from Ref. \cite{ring97}
Our calculation started from the Lagrangian density \cite{rin96, ser92, bog77, sah18, pat91}

\begin{equation}
\begin{split}
L &=\overline{\psi_i}(i\gamma_{\mu}\partial_{\mu}-M)\psi_{i}+\frac{1}{2}\partial^{\mu}\sigma\partial_{\mu}\sigma
-\frac{1}{2}m_{\sigma}^{2}\sigma^{2}-\frac{1}{3}g_{2}\sigma^{3} \\& -\frac{1}{4}g_{3}\sigma^{4}-g_{s}\overline{\psi_{i}}\psi_{i}\sigma
-\frac{1}{4}\Omega^{\mu\nu}\Omega_{\mu\nu}+\frac{1}{2}m_{\omega}^{2}V^{\mu}V_{\mu} \\& +\frac{1}{4}c_{3}(V_{\mu}V^{\mu})^{2}-g_{\omega}\overline{\psi_i}\gamma^{\mu}\psi_{i}V_{\mu}
-\frac{1}{4}\overrightarrow{B}^{\mu\nu}\overrightarrow{B}_{\mu\nu} \\& +\frac{1}{2}m_{\rho}^{2}\overrightarrow{R}^{\mu}.\overrightarrow{R}_{\mu}-g_{\rho}\overline{\psi_i}\gamma^{\mu}\overrightarrow{\tau}\psi_{i}\overrightarrow{R}^{\mu}
 \\& -\frac{1}{4}F^{\mu\nu}F_{\mu\nu}-e\overline{\psi_i}\gamma^{\mu}\frac{(1-\tau_{3i})}{2}\psi_{i}A_{\mu}
\end{split}
\end{equation}

Here $\sigma$, $V_{\mu}$, $\overrightarrow{R}^{\mu}$, represent the fields for  isoscalar-scalar meson $\sigma$, isoscalar-vector meson $\omega$ and isovector-vector meson $\rho$  respectively. $A^{\mu}$ represents the electromagnetic field. In this model the nucleons are considered as Dirac spinors represented as $\psi$ in the above equations. $g_{s}$, $g_{\omega}$, $g_{\rho}$ and $\frac{e^{2}}{4\pi}$ represents the coupling constants for $\sigma$, $\omega$, $\rho$ mesons and photon respectively.
The field tensors for the vector mesons and the electromagnetic field are given below
\begin{center}

$\Omega^{\mu\nu}=\partial^{\mu}V^{\nu}-\partial^{\nu}V^{\mu}$

$\overrightarrow{B}^{\mu\nu}=\partial^{\mu}\overrightarrow{R}^{\nu}-\partial^{\nu}\overrightarrow{R}^{\mu}-g_{\rho}(\overrightarrow{R}^{\mu}\times \overrightarrow{R}^{\nu})$

$F^{\mu\nu}=\partial^{\mu}A^{\nu}-\partial^{\nu}A^{\mu}$
\end{center}
The field equations for nucleons and mesons i.e Klein Gordan equation for mesons and Dirac equation for nucleons are obtained from the above Lagrangian density 
by using classical variational principle.
 Here we neglect the contribution of antiparticles i.e in another way we can say that we are not  considering the negative energy solutions of Dirac equation.
The static solutions of the field equations gives us the ground state properties of a nucleus.
Due to this the meson and electro magnetic field are time independent.
Also due to time reversal symmetry the spatial parts of vector potential (V) and $\rho$ and the electromagnetic potential (A) vanish. 
But while dealing with odd odd nuclei or odd A nuclei the time reversal symmetry of the mean field breaks. 
Hence the space like components do not vanish any more.
 So, to deal with such difficulty we use blocking approximation where one pair of conjugate states $\pm m$ is taken out of the pairing scheme. The odd particle remains in one of the states leaving the conjugate state empty. By blocking different states around the fermi level one finds the ground state energy of the odd nucleus. In case of odd odd nuclei the blocking process is done for both odd nucleon. The blocking restores the time reversal symmetry.
 The spinors are the eigen vectors of the static Dirac equation, which then produces the single particle energies as eigen values.
 The wave functions are expanded in a deformed harmonic oscillator basis. 
The maximum oscillator shells for both bosons and fermions are taken as 20. The details regarding the expansion method can be seen \cite{gam90, pat91}.
 The solutions are carried out by a self consistent iteration method with initial deformation value \cite{ser92, bog77, del01}
 $\beta_{0}=-0.3, -0.2, -0.1, 0.0, 0.1, 0.2, 0.3$.
 Starting with values for potentials (scalar and vector potential) (chosen by guess) one can solve Dirac equation for the spinors $\psi$. These are then used to calculate the densities $\rho_{s}$ , $\rho_{v}$, $\rho_{\rho}$ and $\rho_{c}$ which are acting as the source terms for stationary field equations for mesons. From these the meson fields are calculated which then helps in calculating the scalar potential and the vector potential. This repetition is continued until convergence is achieved. 
 By taking these initial deformation values we obtain the ground state solutions. 
The center of mass correction energy is given by
	$E_{c.m}=\frac{3}{4}(41A^{\frac{-1}{3}})$.
 Where A represents the mass number of a nucleus.
 The quadrupole deformation parameter $\beta_{2}$ is calculated from the formula	$Q=Q_{n}+Q_{p}=\sqrt{\frac{16\pi}{5}}(\frac{3}{4\pi}AR^{2}\beta_{2})$ 

The matter radius of a given nucleus is given as
	$<r_{m}^{2}>=\frac{1}{A}\int{\rho(r_{\perp},z)r^{2}d\tau}$
 where $\rho(r_{\perp},z)$ represents the axially deformed density. 
The total energy of the system can be obtained as 
$E_{total}=E_{part}+E_{\sigma}+E_{\omega}+E_{\rho}+E_{C}+E_{pair}+E_{c.m}$ 
. In this case a constant gap approximation for proton
	$\Delta_{p}=\frac{RB_{s}e^{SI-tI^{2}}}{Z^{\frac{1}{3}}}$
and neutron  $\Delta_{n}=\frac{RB_{s}e^{-SI-tI^{2}}}{A^{\frac{1}{3}}}$ 
with RMF-BCS pairing effect \cite{gam90, mad88, wer94} is taken into account and R = 5.72, S = 0.118, t = 8.12, $B_S$ = 1 and $I = \frac{(N-Z)}{(N+Z)}$
are considered during calculation.
	
\subsection{The Universal Decay Law}
A relationship is established between the Q-values of the departing particles and the monopole radioactive decay half-life periods by the Universal decay law. Additionally, it establishes a connection between the masses and charges that participate in the decay process and the decay half-life. The formula is equally applicable for the calculation of cluster decay half-live as well as alpha decay half-life. The formula is popularly known as Universal decay law (UDL) \cite{ism17} given as follows;  

\begin{equation}
Log_{10}T_{1/2}=aZ_{c}Z_{d}\sqrt{\frac{\mu}{E}}+b\sqrt{\mu Z_{c}Z_{d}(A_{d}^{1/3}+A_{c}^{1/3})}+c
\end{equation}
$\mu=\frac{A_{d}A_{c}}{(A_{d}+A_{c})}$
Where $A_{d}$, $Z_{d}$ and $A_{c}$, $Z_{c}$ represents the mass number and atomic number of the daughter nucleus and emitted cluster respectively, 
Where a, b, and c are constants taken from \cite{ism17}.
 We have applied the above UDL formula to calculate cluster decay half-life periods.

\subsubsection{\bf{Modified Universal Decay Law}}

The addition of asymmetry term to Universal decay law gives rise to the modified universal decay law(MUDL) \cite{akra19}.
 The MUDL formula is as follows;
\begin{equation}
\begin{split}
	Log_{10}T_{1/2}(\alpha) &=aZ_{\alpha}Z_{d}\sqrt{\frac{A}{Q_{\alpha}}} \\& +b\sqrt{AZ_{\alpha}Z_{d}(A_{d}^{1/3}+A_{\alpha}^{1/3})}+c+dI+eI^{2}
\end{split}
\end{equation}
 Here the asymmetry term is $I=\frac{N-Z}{A}$. 
 and $A=\frac{A_{d}A_{\alpha}}{(A_{d}+A_{\alpha})}$
Here $A_{d}$ and $A_{\alpha}$ represent the mass number of daughter nuclei and alpha particle respectively.
 a, b, c, d and e are the fitting parameters determined by fitting to 365 experimental values.
 In our case we take these values from reference \cite{akra19}.
 We have applied the MUDL formula to calculate $\alpha$-decay half-life periods.
\subsection{Akre Formula}

To calculate $\alpha$-decay half-life periods Royer proposed a semi-empirical formula \cite{roy00}. It shows a relation between the $Log_{10}T_{1/2}(\alpha)$ with charge number (Z) and mass number (A) of parent nuclei as well as the energy released (Q) during the emission of $\alpha$ particle. The relation is as follows
\begin{equation}
Log_{10}T_{1/2}(\alpha)=a+bA^{1/6}\sqrt{Z}+\frac{cZ}{\sqrt{Q_{\alpha}}}
\end{equation}
 Where a, b, c are constants and obtained from experimental fitting.
Akrawy and Poenaru modified the Royer formula by including the isospin asymmetry term in Royer formula \cite{akr19}.
The formula is as follows;
\begin{equation}
Log_{10}T_{1/2}(\alpha)=a+bA^{1/6}\sqrt{Z}+\frac{cZ}{\sqrt{Q_{\alpha}}}+dI+eI^{2}
\end{equation}
 Where Z, N, A represents atomic number, neutron number and mass number of parent nucleus respectively. 
Here a, b, c, d and e are constants obtained by fitting to experimental values. 
$I=\frac{(N-Z)}{A}$ represents the asymmetry term. 
The parameters are taken from the reference \cite{akr19}.
\subsection{The Scaling law of Horoi}

This is the first model independent law for all known cluster decay phenomena. 
The empirical formula was introduced by Horoi et al \cite{hor04}. 
To determine the cluster decay half-lives of Es isotopes we use the formula \cite{ade16} given below.
 \begin{equation}
Log_{10}T_{1/2}=(a_{1}\sqrt{\mu}+b_{1})[\frac{(Z_{c}Z_{d})^{y}}{\sqrt{Q}}-7]+(a_{2}\sqrt{\mu}+b_{2})
\end{equation}
Here $\mu=\frac{A_{c}A_{d}}{(A_{c}+A_{d})}$
Where $A_{d}$, $Z_{d}$ and $A_{c}$, $Z_{c}$ represents the mass number and atomic number of daughter nucleus and emitted cluster respectively.
The constant parameters $a_1$, $b_1$, $a_2$, $b_2$ and y are taken from \cite{ade16}. These are $a_{1}=6.8\pm 0.2$, $b_{1}=-7.5\pm 0.5$, $a_{2}=6.9\pm 1.4$, $b_{2}=-22.4\pm 2.0$ and $y=0.607\pm 0.004$.
\subsection{Beta Decay}
A new semi-empirical formula for $\beta$-decay half-life has been introduced by Hadi \textit{et al.} \cite{sob23}. 
The formula is unique because it remains same for all types of beta decay and needs only the information of $Q_{\beta}$-values. 
The formula is given below;
\begin{equation}
Log_{10}T_{1/2}=a_{Z}Z+a_{A}A+a_{Q}Q^{\frac{-1}{4}}+a_{I}I+a
\end{equation}
Where Z and A represents the atomic number and mass number of parent nucleus respectively. 
I is the asymmetry term and is equal to $I=\frac{N-Z}{N+Z}$=$\frac{N-Z}{A}$.
$a_{Z}$, $a_{A}$, $a_{Q}$, $a_{I}$ and $a$ are five real constants used in the above formula. 
These constants are obtained separately from experimental data for different types of beta decay. These are taken from the article \cite{sob23}.

The $\beta^-$ decay energy $Q_{\beta}$-value has been calculated using the corresponding B.E. values in the equation given below; 
\begin{center}
$Q_{\beta^-}=B.E(Z+1,N-1)-B.E(Z,N)+(m_{n}-m_{H})c^{2}$
\end{center}
Here $(m_{n}-m_{H})c^{2} \simeq 0.782 MeV$ represents the difference in masses of neutron and H-atom. 
Similarly for $\beta^+$ decay energy $Q_{\beta}$-value has been calculated using the corresponding B.E. values in the equation given below; 
\begin{center}
$Q_{\beta^+} = B.E(Z-1,N+1)-B.E(Z,N)+(m_{H}-m_{n})c^{2}-2m_{e}c^{2}$
\end{center}
In terms of $Q_{\beta^{+}}$, we have calculated the Q-value for electron capture using the relation below;  
\begin{center}
$Q_{EC} = Q_{\beta^{+}}+2m_{e}c^{2}-$ B.E. of K-shell electron
\end{center}
For Einsteinium the B.E. of the K-shell electron is $\approx$ 0.125 MeV.
So,
\begin{center}
$Q_{EC} = Q_{\beta^{+}}+0.897$
\end{center}

\section{Results and discussion}
In this work we examine the ground state properties such as binding energy(B.E.) binding energy per nucleon(B.E./A), one neutron separation energy ($S_{1n}$), two neutron separation energy ($S_{2n}$), differential variation of two neutron separation energy ($dS_{2n}$) and the quadrupole deformation parameter ($\beta_{2}$), neutron skin thickness ($r_{np}$) and charge radius($r_{c}$) of $^{240-259}$Es$_{99}$ isotopes. The single particle energies for $^{247}Es$ and $^{257}Es$ isotopes are also investigated along with its variation with deformation parameter by drawing the Nilson plot for $^{257}Es$ isotope.
 
We have calculated the $\alpha$-decay half-lives using experimentally available $Q_{\alpha}$ values obtained from National Nuclear Data Center(NNDC)\cite{nndc} and our calculated $Q_{\alpha}$ values in two semi-empirical formulas such as MUDL \cite{akra19} and AKRE \cite{akr19}. 
Not only that we have also calculated the cluster decay half life for $^{8}$Be$_{4}$, 
$^{12}$C$_{6}$, $^{14}$C$_{6}$, $^{16}$O$_{8}$
decay using our calculated Q-values in Horoi \cite{ade16} formula and in UDL \cite{ism17}. 

We have also calculated the $\beta$-decay half-lives using a semi-empirical formula from the article \cite{sob23} 
for the above mentioned Es isotopes to analyze the favored decay modes among them.
	
	Among so many force parameter set, we choose two set of force parameters that are 
NL3* \cite{lol09} and NL-SH \cite{sha93} 
while calculating the bulk properties. A detailed of the parameters is mentioned in the table given below;
%\begin{table}
\begin{center}
Table-1\\
The force parameters of the RMF model\\
\vspace{0.5 cm}
\begin{tabular}{||c||c||c||}
\hline
Parameters & NL3* & NL-SH \\
\hline
%\hhline{|=|=|=|=|}
$M_{n}$ & 939.0 & 939.0 \\
\hline
$m_{\sigma}$ & 502.5742 & 526.059\\
\hline
$m_{\omega}$ & 782.600 &  783.0 \\
\hline
$m_{\rho}$ & 763.000 & 763.0 \\
\hline
$g_{\sigma}$ & 10.0944 & 10.444 \\
\hline
$g_{\omega}$ & 12.8065 & 12.945 \\
\hline
$g_{\rho}$ & 4.5748 & 4.383 \\
\hline
$g_{2}$ & -10.8093 & -6.9099\\
\hline
$g_{3}$ & -30.1486 & -15.8337\\
\hline
%\hhline{|=|=|=|=|}
\end{tabular}
\end{center}
%\end{table}
Due to the differences in the values of coupling constants and masses of mesons, NL3* and NL-SH parameter sets some times produce different results. So, we call RMF model a parameter dependent model.  
We take maximum oscillator shell $N_F$=$N_B$=20 both for fermions and bosons during numerical computation.
	
	 The binding energy (B.E.) is one of the important structural property of nuclei, which helps in predicting the validity of nuclear models. Fig.\ref{fig1} shows the variation of  B.E. values as a function of neutron number for NL3*, NL-SH and Finite Range Droplet Model (FRDM) \cite{mol16}. Each curve shows similar trend with enhanced values for even neutron numbers.
The B.E. values obtained for NL3* matches well with that of FRDM.  
The enhanced values of binding energy for even N may be due to the pairing effect of neutron.
\vspace{0.5cm}
\begin{center}
%\vspace{0.3cm}
\includegraphics[width=8cm, angle=0]{beesf.eps}
\figcaption{\label{fig1}   Variation of binding energy with neutron number of Es,  estimated for RMF model with NL3* and NL-SH parameter set and compared with FRDM \cite{mol16}. }
\end{center}
 \vspace{0.5cm}	
 	While discussing about the stability of a nucleus, B.E./A. is crucial. We compare our computed B.E./A values in Fig.\ref{fig2} with the experimentally accessible data. 
 All the three curves have a similar trait. The obtained B.E./A values for the NL3* parameter exhibit a strong correlation with the values obtained from experiments. For a given N, the NL-SH curve exhibits higher B.E./A values than the NL3* force parameters. 
At N = 142 and 144, we get the highest B.E./A values of 7.493 MeV and 7.525 MeV, respectively, for the NL3* and NL-SH parameter set.
Experimentally it is maximum at N = 144 with a value of 7.486 MeV.
Based on the analysis of B.E./A, we find that $^{243}$Es$_{99}$ is the most stable isotope in the series. 
%\vspace{1.0cm}
\begin{center}
\vspace{0.5cm}
\includegraphics[width=8cm, angle=0]{beaesf.eps}
\figcaption{\label{fig2} Variation of B.E./A with neutron number of Es, estimated for RMF model with NL3* and NL-SH parameter set and compared with experimental values obtained from National Nuclear Data Centre (NNDC).}
\end{center}
\vspace{0.5cm}
We have also evaluated the standard deviations ($\sigma$) in B.E./A for NL3* and NL-SH parameter in comparison to the experimental data using an analogous equation mentioned in the article \cite{san17} 
The equation is as follows;
\begin{equation}
\sigma = \bigg\{\frac{1}{n-1}\sum_{i=1}^{n}\bigg(\frac{B.E.}{A}^{calc}-\frac{B.E.}{A}^{exp}\bigg)^{2}\bigg\}^{1/2} \nonumber
\end{equation}
The standard deviations is found to be 0.00704 and 0.03158 for NL3* and NL-SH respectively.

	Proton and neutron distributions in a nucleus are measured in terms of proton and neutron radius respectively. These are actually the key to determine the size of the nuclear system. These observables are directly related to the bulk properties of nuclear matter but can also be related with the nature of nuclear interactions. In this context, another nuclear observable named neutron skin thickness which is related to nucleon density distribution can be considered as important because it is sensitive towards the nuclear surface properties. In nuclei with large neutron excess, the neutron density is expected to extend beyond the proton density which in turn gives rise to neutron skin \cite{ren16, hag15}. 
Fig.\ref{fig3} represents the variation of skin thickness with neutron number for both parameter set. We observe that with increase in neutron number the skin thickness increases reflecting the pressure of symmetry energy.
\begin{center}
\vspace{0.5cm}
\includegraphics[width=8cm, angle=0]{skinEsf.eps}
\figcaption{\label{fig3}  Variation of skin thickness with neutron number of Es, estimated for RMF model with NL3* and NL-SH parameter set}
\end{center}
\vspace{0.5cm}
\begin{center}
\vspace{0.5cm}
\includegraphics[width=8cm, angle=0]{rcEsf.eps}
\figcaption{\label{fig4} Variation of r$_{c}$ as a function of neutron number of Es, estimated for RMF model with NL3* and NL-SH parameter set}
\end{center}
\vspace{0.5cm}
	In an isotopic series, nuclear charge radii ($r_{c}$) helps in the search of shell effects, because they are sensitive towards the changes in nuclear deformation and nuclear size. Some times a prominent kink is observed across spherical shell closures \cite{kre14, gor19, rep21, day21}. 
Around N = 40 sub-shell closure, $r_{c}$ shows a localized effect for Nickel isotope in laser spectroscopy measurements \cite{war24} 
relative to droplet model\cite{mal22, yan23}. 
While at N = 32 sub-shell closure, in neuron rich Potassium isotope \cite{kos21} 
the charge radius doesn't manifest itself. The experimental information regarding charge radius around the shell gaps in heaviest actinide nuclei and beyond is limited because of their  production capabilities. The experiments based on combination of highly sensitive laser spectroscopy techniques with multiple production schemes were conducted by Jessica Warbinek et al \cite{war24}. They also suggested that the
weak shell effects in this region do not influence the charge radii \cite{war24}. 
$r_{c}$ was also observed of not showing any kink at N = 152 in our earlier work of Fm \cite {das25}. 
So, to check the manifestation of $r_{c}$ around N = 154 gap which is reflected in separation energies and single particle energy levels, we have plotted $r_{c}$ verses neutron number in Fig.\ref{fig4}. The curves don't display any special effect at N = 154, rather the curves display odd even staggering (OES).
The shell structure along with many body correlations provide a greater impact on the charge radius which in turn results in the local fluctuations including the odd even staggering (OES) \cite{gro20}.
The polarisation effects of an odd nucleus in a particular shell model (or generally known as one quasi particle) orbital also contributes towards OES \cite{gro20}.
In some semi-magic isotopic chains of spherical nuclei the self consistent coupling between the neutron pairing field and the proton density enlighten us with the knowledge of OES of charge radii \cite{gro20}. 
 In our case, the data also exhibit features characteristic of odd-even staggering
(OES) in nuclear radii. Both the curves display oscillatory behavior, with
elevated values observed for nuclei with odd neutron numbers, as shown in Fig.\ref{fig4}. This trend arises because, compared to their neighboring even-neutron
isotopes (illustrated in Fig. 5), odd-neutron nuclei tend to exhibit greater
deformation. This is due to the unpaired neutron occupying specific deformed
orbitals, leading to a staggering effect in the nuclear radius.’ 

 	The shape of a nucleus depends both on macroscopic bulk properties and also on microscopic properties like shell effect. 
Nucleus with partially filled Shells with nucleons, the valence nucleons tend to polarize the core and deforms the mass of the nucleus. We can describe the deformation of a nucleus by multi pole expansion, with the quadrupole deformation being the important parameter towards the determination of the deviation from the
spherical shape.
 Such quadrupole shapes may either have axial symmetry with prolate or oblate shape or triaxial shape. In some region of periodic table the shapes play very crucial role towards the structural study of a nucleus.
The shape may change due to the change in proton and neutron numbers or with excitation energy or angular momentum within the same nucleus. The changes may be due to rearrangement of the orbital configuration of the nucleons or due to dynamic response of the nuclear system to rotation. 
We plot quadrupole deformation parameter ($\beta_{2}$) as a function of neutron number of Es both for NL3* and NL-SH force parameter in Fig.\ref{fig5}. 
\vspace{0.5cm}
\begin{center}
\includegraphics[width=8cm]{bEsf.eps}
\figcaption{\label{fig5}   Variation of $\beta_{2}$ with neutron number of Es, estimated for RMF model with NL3* and NL-SH parameter set }
\end{center}
%\vspace{-0.5cm}
Both the NL3* and NL-SH curves show prolate shape through out the isotopic series. The odd even staggering is also seen in this Figure. For NL-SH parameter we get maximum deformation at N = 141 but for NL3* it occurs at N = 149. After that $\beta_{2}$ decreases with increase in neutron number. At N = 154 we get $\beta_{2}$ = 0.28 for NL-SH parameter and 0.2989 for NL3* parameter.

	To understand nuclear structure far from stability line, it is important to have the knowledge about the energy difference between adjacent shells which in turn helps in identification of shell and sub shell closures.
A large shell gap is a sign of shell or sub-shell closure.
As a result of nucleon pairing, the position of the two neutron drip line may diverge from single neutron drip line. 
We thus computed both.
We have calculated $S_{1n}$ and $S_{2n}$ substituting B.E. values obtained for both NL3* and NL-SH parameter set in the formulas given below;
\begin{center}
$S_{1n}=B.E.(N,Z)-B.E.(N-1,Z)$	
$S_{2n}=B.E.(N,Z)-B.E.(N-2,Z)$
\end{center}
The variation of $S_{1n}$ and $S_{2n}$ as a function of neutron number of Es is shown in Fig.\ref{fig6} and Fig.\ref{fig7} respectively.
\newpage
\vspace{0.5cm}
%\begin{figure}[h!]
\begin{center}
\includegraphics[width=8cm, angle=0]{s1nesf.eps}
\figcaption{\label{fig6}   Variation of $S_{1n}$ with neutron number of Es, estimated for RMF model with NL3* and NL-SH parameter set and compared with FRDM values \cite{mol16} and experimental values obtained from National Nuclear Data Centre (NNDC)}
%\vspace{0.5cm}
\end{center}
%\end{figure}
\vspace{0.5cm}
\begin{center}
\includegraphics[width=8cm, angle=0]{s2nesf.eps}
\figcaption{\label{fig7}   Variation of $S_{2n}$ with neutron number of Es, estimated for RMF model with NL3* and NL-SH parameter set and compared with FRDM values \cite{mol16} and experimental values obtained from National Nuclear Data Centre (NNDC) }
\vspace{0.5cm}
\end{center}
Not only that we have also the differential variation of two neutron separation energies ($dS_{2n}$) using $S_{2n}$ values in the formula given below;
\begin{center}
	$dS_{2n}=\frac{S_{2n}(N+2,Z)-S_{2n}(N,Z)}{2}$
\end{center}
Fig.\ref{fig8} depicts the change in $dS_{2n}$(N,Z) with increasing neutron number of Es.
We have also compared our estimated results of $S_{1n}$,  $S_{2n}$ and  $dS_{2n}$  with experimentally obtained values \cite{nndc} and with that of the values obtained from  FRDM \cite{mol16}. 
Each of the four curves in Fig.\ref{fig6} possess oscillating behavior with peaks at even neutron counts. Keeping the trend same, our calculated $S_{1n}$ values show some divergence from the experimental values. The divergence in the Relativistic Mean Field Formalism is explained in detail in \cite{jos22}. Some times better results may be produced by introducing Covariant Density Functional Theory (CDFT)\cite{els22}.
We present our results upto N = 160 because we are getting the drip line for the RMF-NL3* curve at N = 163 and the negative value for one neutron separation energy for the RMF-NL-SH curve at N = 161. 
We thus predict the neutron drip line to be located close to that neutron value.
 In Fig.\ref{fig7} all the four curves are observed of displaying similar behavior. $S_{2n}$ decreases with increasing neutron number with two sharp drops at N = 148 and N = 154 
 for both parameter set. These drops indicate a greater stability of $^{247}$Es$_{99}$ and $^{253}$Es$_{99}$ isotopes against neutron separation indicating a possibility of shell/sub-shell closure at N = 148 and N = 154. 
\vspace{0.7cm}
\begin{center}
\includegraphics[width=8cm, angle=0]{ds2nesf.eps}
\figcaption{\label{fig8}   Variation of $dS_{2n}$ with neutron number of Es, estimated for RMF model with NL3* and NL-SH parameter set and compared with FRDM values \cite{mol16} and experimental values}.
\vspace{0.3cm}
\end{center}

	In Fig.\ref{fig8} both NL3* and NL-SH curves show deep at N = 154 which is in accordance with the result obtained from Fig.\ref{fig7}. This indicates a possible shell/sub-shell closure at N = 154. Apart from that a deep is observed at N = 148 for NL3* parameter. Except this we get deeps at N = 147, 156, 159 for NL-SH curve and deeps at N = 151 and 157 for NL3* curve. 
But the curves corresponding to FRDM \cite{mol16} and experimental \cite{nndc} results show deep at N = 152. 

	The neutron single particle energy levels for $^{247}Es$ and $^{257}Es$ isotopes for zero quadrupole deformation are plotted in Fig.\ref{fig9}. The gap between the levels are also marked here. Where we can see large gaps between levels $3p_{1/2}$ and $1i_{11/2}$ that is for N = 126, between $1i_{11/2}$ and $1j_{15/2}$ that is N = 138, $1j_{15/2}$ and $2g_{9/2}$ that is N = 154 and in between $2g_{9/2}$ and $2g_{7/2}$ that is N = 164. 
We also observe a gap at N = 138 for our RMF model similar to Ummukulsu \textit{et al} \cite{umm23}. Also at N = 164 a spherical shell closure is observed similar to the earlier prediction of \cite{bre94, gus67, isi07}.

We have also plotted the variation of single particle energies against quadrupole deformation parameter ($\beta_{2}$) in Fig.\ref{fig10}. Here also the gap at N = 154 is clearly observed in between $\frac{9}{2}^{+}[6 1 5]$ and $\frac{11}{2}^{-}[7 2 5]$ Nilson quantum levels. Rather the gap is slightly larger for deformed shells compared to zero deformation.       
\vspace{0.5cm}
\begin{center}
\includegraphics[width=8cm, angle=0]{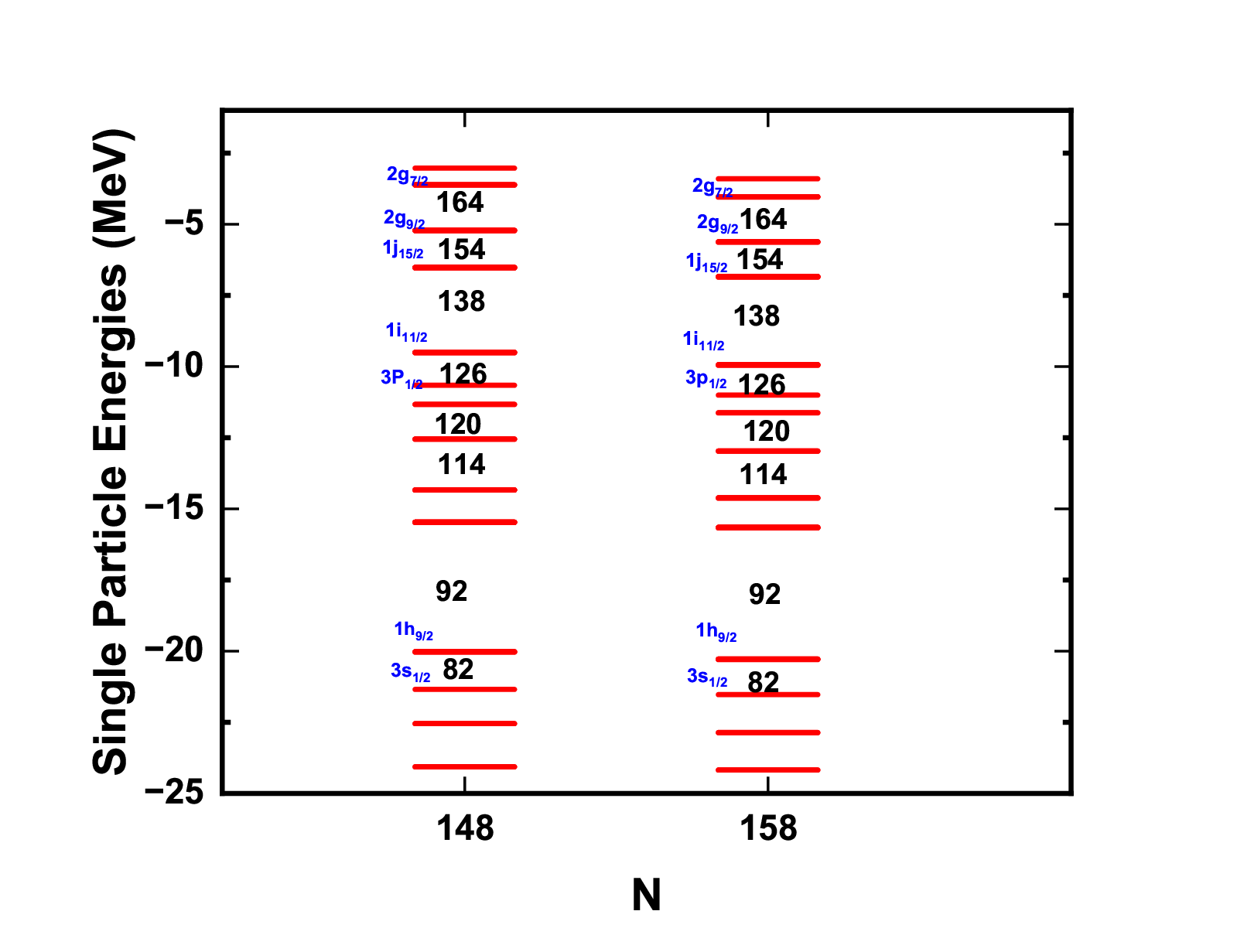}
\figcaption{\label{fig9}   Single particle energy levels of $^{247, 257}Es$ isotopes,  estimated for RMF model with NL-SH parameter set.}
\vspace{0.5cm}
\end{center}

\vspace{0.5cm}
\begin{center}
\includegraphics[width=8cm, angle=0]{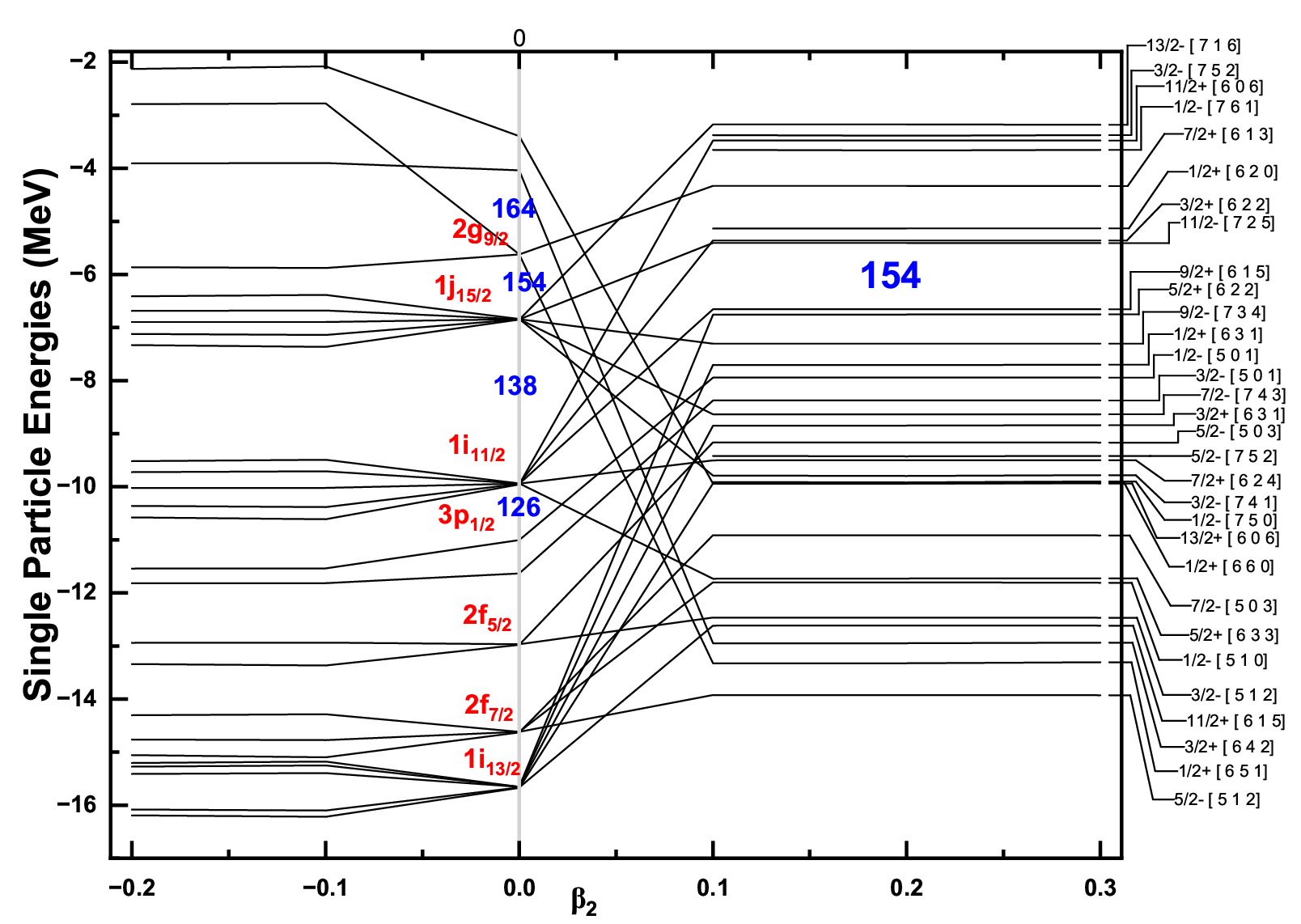}
\figcaption{\label{fig10}  The single-particle energies of
$^{257}Es$, as a function of the deformation parameter ($\beta_{2}$), estimated for RMF model with NL-SH parameter set.}
\vspace{0.5cm}
\end{center}	
	
The decay energy is also an important aspect in the investigation of shell closure. Minimum Q-values are manifested at the shell stabilized neutron numbers when its variation is plotted against parent neutron number. The alpha decay Q-values are calculated for both NL3* and NL-SH parameter set and are compared with that of experimental values and FRDM values with an error assessment as a form of standard deviation from experimental values using a formula \cite{san17}
given below;
\begin{equation}
\sigma = \bigg\{\frac{1}{n-1}\sum_{i=1}^{n}\bigg(Q_{\alpha}^{calc}-Q_{\alpha}^{exp}\bigg)^{2}\bigg\}^{1/2} \nonumber
\end{equation} 

The $Q_{\alpha}$ values are estimated using B.E. values in the equation below;
\begin{center}
$Q(decay)= B.E.(Daughter)+B.E.(cluster)-B.E.(Es)$
\end{center}
The B.E. of $\alpha$ particle is taken as 28.3 MeV.
%\vspace{0.5cm}

The standard deviations in $Q_{\alpha}$ are found to be 0.40172, 0.68979 and 0.30853 for NL3* , NL-SH and FRDM respectively.
\vspace{0.7cm}
\begin{center}
\includegraphics[width=8cm, angle=0]{qalphaesf.eps}
\figcaption{\label{fig11}   Variation of $\alpha$-decay energy ($Q_{\alpha}$) with neutron number of Es, estimated for RMF model with NL3* and NL-SH parameter set and compared with FRDM values \cite{mol16} and experimental values obtained from National Nuclear Data Centre (NNDC) }
\vspace{0.3cm}
\end{center}
 
In our case, a clear minimum is observed at N = 154 for NL-SH parameter set in Fig.\ref{fig11}. So, shell/sub-shell closure is expected at N = 154.

	The $\alpha$-decay half-lives ($Log_{10}T_{1/2}(\alpha)$) versus parent neutron number are plotted in Fig.\ref{fig12} and Fig.\ref{fig13} for NL3* and NL-SH parameter set respectively.
Both calculated and experimentally available Q-values are used in MUDL \cite{akra19} 
and AKRE \cite{akr19} formula for half-life calculation. 
A greater value of $Log_{10}T_{1/2}(\alpha)$ indicates a shell stabilized parent nucleus.
i.e Increasing values of $Log_{10}T_{1/2}(\alpha)$ is an indication of the increasing stability of the parent isotopes against $\alpha$-decay. 
For both NL3* and NL-SH parameter we get a kink at N = 148 which indicates the stability of $^{247}$Es$_{99}$ isotope against $\alpha$-decay. Except the kink, we get peaks at N = 150, 152, 154 while calculating $\alpha$-decay for Q-values obtained for NL-SH parameter set. This indicates the stability of $^{249}$Es$_{99}$, $^{251}$Es$_{99}$, $^{253}$Es$_{99}$ isotopes against $\alpha$-decay. While calculating $Log_{10}T_{1/2}(\alpha)$ for experimental Q-values we get a small kink at N = 152 which can be seen in Fig.\ref{fig13}. Except these peaks we get a deep at N = 143 for MUDL-NL-SH and AKRE-NL-SH curve indicating less stability of $^{242}Es$ against $\alpha$-decay. 
\vspace{0.7cm}
\begin{center}
\includegraphics[width=8cm, angle=0]{talEsn3f.eps}
\figcaption{\label{fig12}   Variation of $Log_{10}T_{1/2}(\alpha)$ with neutron number of Es for NL3* parameter. }
\vspace{0.3cm}
\end{center}
\vspace{0.5cm}
\begin{center}
\includegraphics[width=7cm, angle=0]{talEsnsf.eps}
\figcaption{\label{fig13}   Variation of $Log_{10}T_{1/2}(\alpha)$ with neutron number of Es for NL-SH parameter. }
\vspace{0.5cm}
\end{center}

	Away from the stability line, the $\beta$-decay processes play an important role.
To find the favorable decay mode for Es isotopes in the $^{240-259}$Es$_{99}$ isotopic range, we have compared the $\alpha$-decay half-lives with $\beta$-decay half-lives in Table-2. Where $^{240, 241, 242, 253, 259}Es$ isotopes are found to possess $\alpha$- decay as their dominant mode of decay. $^{243-250}Es$ isotopes possess $\beta^{+}$-decay as their dominant decay mode. It is observed that electron capture is the dominating decay mode for $^{251}Es$ isotope. There is an equal probability of electron capture and positive beta decay for $^{252}Es$ isotopes. Apart from that $^{254-258}Es$ isotopes are found to decay via emitting $\beta^{-}$ particle. The decay modes are very well reproduced. They show excellent match with the experimental decay modes. 

\vspace{0.7cm}
\begin{center}
\includegraphics[width=8cm, angle=0]{tEsn3f.eps}
\figcaption{\label{fig14}   Variation of $Log_{10}T_{1/2}$ with parent (Es) neutron number for NL3* parameter. }
\vspace{0.3cm}
\end{center}

	Now a days cluster decay is drawing the attention of many nuclear structural investigators because it helps in analysing the shell structure of a nucleus. A detailed investigation of cluster decay for both ground and intrinsic excited states of $^{112–122} Ba$ isotopes has been carried out by Joshua T. Majekodunmi et al \cite{maj22}. Detailed description in recent advancement in cluster decay can also be observed in \cite{hec22, ism25}. We have also estimated cluster decay half-lives using Horoi \cite{ade16} formula and UDL formula \cite{ism17}.
Cluster decay is possible only when the decay energy is positive.
We have plotted cluster decay half-lives ($Log_{10}T_{1/2}$) for $^{8}$Be$_{4}$, 
$^{12}$C$_{6}$, $^{14}$C$_{6}$ and  $^{16}$O$_{8}$ 
decays against parent neutron number (N) For the NL3* force parameter and NL-SH force parameter in Fig.\ref{fig14} and Fig.\ref{fig15} respectively.
In general cluster decay half-life is minimum for those decays which leads to doubly magic daughter nucleus. 

Here the shell structure plays an important role in determining the type of cluster to decay from the parent nucleus. Among $^{8}$Be$_{4}$, 
$^{12}$C$_{6}$, $^{14}$C$_{6}$ and  $^{16}$O$_{8}$ clusters $^{14}C$ is observed to be the favorable cluster to decay in the isotopic chain. For $^{8}$Be$_{4}$ cluster decay, we observe longer half-live for parent isotopes which can be observed in Fig.\ref{fig14} and Fig.\ref{fig15}. In Fig.\ref{fig14} we observe a sudden deep at N = 158 ($^{257}Es$) for $^{8}$Be$_{4}$ cluster decay for the Horoi curve which indicates a possibility of neutron shell closure at N = 154 for $^{249}Am_{95}$ daughter nucleus.
\vspace{0.5cm}
\begin{center}
\includegraphics[width=8cm, angle=0]{tEsnsf.eps}
\figcaption{\label{fig15} Variation of $Log_{10}T_{1/2}$ with parent (Es) neutron number for NL-SH parameter. }
%\vspace{0.5cm}
\end{center}

%\newpage

\end{multicols}
\begin{landscape}
Table-2

 $log_{10}T_{1/2}(\alpha)$, $log_{10}T_{1/2}(\beta^{-})$, $log_{10}T_{1/2}(\beta^{+})$, $log_{10}T_{1/2}(EC)$ values calculated using Q-values obtained for NL3* and NL-SH parameter set and also accessible experimental Q-values obtained from NNDC \cite{nndc}%$log_{10}T_{1/2}(SF)$ values obtained  from \cite{zho05} 
 of $^{240-259}$Es$_{99}$ isotopes. 
%\begin{table*}
\begin{center}
\begin{tabular}{|c|c|c|c|c|c|c|c|c|c|c|c|c|c|c|c|}
\hline
 & & \multicolumn{4}{c|}{$log_{10}T_{1/2}(\alpha)$ } & \multicolumn{9}{c|}{$log_{10}T_{1/2}(\beta)$ } &  \\
\cline{3-15}
  &  & \multicolumn{2}{c|}{MUDL} & \multicolumn{2}{c|}{AKRE} & \multicolumn{3}{c|}{$log_{10}T_{1/2}(\beta^{-})$ } & \multicolumn{3}{c|}{$log_{10}T_{1/2}(\beta^{+})$ }& \multicolumn{3}{c|}{$log_{10}T_{1/2}(EC)$ } & \\
\cline{3-15}
A & N & NL3* & NL-SH & NL3* & NL-SH & NL3* & NL-SH & EXP & NL3* & NL-SH & EXP & NL3* & NL-SH & EXP & \makecell{DECAY\\ MODE}\\
\hline
240 & 141 & 0.2759 & 2.8471 & 0.2731 & 2.8182 & - & - & -& 1.4104 & 1.4466 & 1.2533 & 1.3747& 1.3999& 1.2515 & $\alpha$ \\
\hline
241 & 142 & 1.0653 & 2.8285 & 1.0614 & 2.8160 & - & - & - & - &-& 1.5314 & 2.1812 &-& 1.5583 &  $\alpha$ \\
\hline
242 & 143 & 1.4542 & 2.2938 & 1.4397 & 2.2708 & - &- & - & 1.8215 & 1.8579 & 1.6170 & 1.8864 &1.9096& 1.7354 &  $\alpha$ \\
\hline
243 & 144 & 2.1542 & 5.4862 & 2.1468 & 5.4625 & - & - &- & - &-& 1.9293 & 4.3212 &-& 2.0598 &  $\beta^{+}$ \\
\hline
244 & 145 & 4.2416 & 6.7417 & 4.1982 & 6.6726 & - & - &- & 2.5591 & 2.3386 & 1.9955 & 2.4056 &2.4506& 2.2276 & $\beta^{+}$ \\
\hline
245 & 146 & 5.0977 & 7.8277 & 5.0778 & 7.7942 & - & - &- & - &- & 2.3604 & - &-& 2.5838 &  $\beta^{+}$ \\
\hline
246 & 147 & 6.2612 & 7.6492 & 6.1961 & 7.5697 & - & - &- &2.8300 & 3.1670 &  2.3884 & 2.9768 &3.0844& 2.7279 &  $\beta^{+}$ \\
\hline
247 & 148 & 7.2841 & 8.5077 & 7.2555 & 8.4728 & - &- & - & - &- & 2.7711 & - & -&3.0804 &  $\beta^{+}$ \\
\hline
248 & 149 & 4.7052 & 6.8913 & 4.6549 & 6.8182 & - & - &- & - &-& 2.7904 & 3.8870 & 4.1185& 3.2286 & $\beta^{+}$ \\
\hline
249 & 150 & 5.2047 & 8.3584 & 5.1890 & 8.3266 & - &- & - & - &-& 3.6354 & - &-& 3.7109 &  $\beta^{+}$ \\
\hline
250 & 151 & 6.4818 & 7.8042 & 6.4111 & 7.7196 & - &- & - & -&- & 3.3606 & - & -&3.7986 &  $\beta^{+}$ \\
\hline
251 & 152 & 6.9386 & 9.6596 & 6.9167 & 9.6237 & - & - &- & - & - & - &-&-& 4.7663 &  EC\\
\hline
252 & 153 & 7.1731 & 9.7574 & 7.0928 & 9.6499 & - & - & 5.7289 & - &-& 4.4163 &-& -& 4.4181 & $\beta^{+}$/ EC\\
\hline
253 & 154 & 8.9137 & 12.8926 & 8.8845 & 12.8425 & - & - &- &- &- & -& -&-& -& $\alpha$ \\
\hline
254 & 155 & 9.7586 & 10.9988 & 9.6483 & 10.8753 & 5.0288 & 4.4663 & 4.0446 & - & - & - & - &-& 5.1365 &  $\beta^{-}$ \\
\hline
255 & 156 & 11.8224 & 13.0670 & 11.7807 & 13.0188 & - & - & 6.2023 & - & - & - & - &-& -& $\beta^{-}$ \\
\hline
256 & 157 & 13.7323 & 14.4632 & 13.5764 & 14.2996 & 3.4521 & 3.1431 & 3.0865 & - & - & - & - & - &-&  $\beta^{-}$ \\
\hline
257 & 158 & 14.8927 & 15.6514 & 14.8376 & 15.5922 & - & - & 8.9818 & - & - & - & - &-& -& $\beta^{-}$ \\
\hline
258 & 159 & 15.9478 & 19.5844 & 15.7645 & 19.3621 & 2.6041 & 2.4383 & -& - & - & - & -&- & -& $\beta^{-}$ \\
\hline
259 & 160 & 18.1711 & 20.9634 & 18.1013 & 20.8785 & - & - & - & - & - & - & - & -&-& $\alpha$ \\
\hline

\end{tabular}
\end{center}
\vspace{0.5cm}
\end{landscape}
%\end{table*}
%\begin{multicols}{2}
%%%%%%%%%%%%%%%%%%%%%%%%%%%%%%%%%%%%%55

\begin{multicols}{2}

\section{Conclusions}
In this article, the nuclear structural properties and different decay modes related to $\alpha$, $\beta$ and cluster-decay were throughly investigated for $^{240-259}$Es$_{99}$ isotopes within the framework of Relativistic Mean Field Model with NL3* and NL-SH parameter set. The ground state properties such as B.E., B.E./A, skin thickness, nuclear charge radius, quadrupole deformation parameter, separation energies and single particle energies etc were estimated and analyzed. These values were also compared with experimentally accessible values.   
Based on the analysis of B.E./A, we find that $^{243}Es_{99}$ is the most stable isotope in the series.
 $^{247}$Es$_{99}$ (N = 148) and $^{253}$Es$_{99}$ (N = 154) isotopes were observed to possess larger separation energies indicating greater stability against neutron separation for both NL-SH and NL3* parameter set.
All Es isotopes were found to exist in prolate shape in their  ground state. 
From the analysis of single particle energies, large shell gaps were observed at N = 126, 138, 154 and 164. From Nilson plot also, a large gap was observed at N = 154. 
 $Q_{\alpha}$ is also found to manifest itself as a minimum for $^{253}$Es$_{99}$ with N = 154 for NL-SH parameter set. For $^{247}$Es$_{99}$ (N = 148) isotope also we get a deep for both parameter sets.

 From the analysis of $\alpha$-decay half lives, $^{247}$Es$_{99}$, $^{251}$Es$_{99}$ and $^{253}$Es$_{99}$ isotopes are found to possess greater stability against $\alpha$ particle emission. Hence we expect a shell/sub-shell closure for neutron at N = 154. 
 A sudden deep at N = 158 ($^{257}Es$) for $^{8}$Be$_{4}$ cluster decay indicates a possibility of neutron shell closure at N = 154 for $^{249}Am_{95}$ daughter nucleus.
 Hence our result needs further experimental investigation. 
  To ascertain the most suitable decay modes and the stability of the specified isotopic series of Es, the $\beta$-decay half-lives are also estimated. The analysis that is now being done on the nuclear structure and decay sensitivity of Es-isotopes might prove useful for future analyses. \\

\end{multicols}

\clearpage
%\end{CJK*}

\end{document}